\documentclass[prd, twocolumn, nofootinbib, floatfix]{revtex4-1}

\usepackage{amsmath}
\usepackage{graphicx}
\usepackage{dcolumn}
\usepackage{bm}
\usepackage{epsfig}
\usepackage{amssymb,latexsym,mathrsfs}
\usepackage{graphicx}
\usepackage{color}
\usepackage{hyperref}

\hypersetup{
    colorlinks=true,
    linkcolor=red,
    citecolor=blue,
} 

\newcommand{\be}{\begin{equation}}
\newcommand{\ee}{\end{equation}}
\newcommand{\bs}{\begin{split}} 
\newcommand{\bea}{\begin{eqnarray}}
\newcommand{\eea}{\end{eqnarray}}

\newcommand{\om}{\Omega_m}

\newcommand{\gl}{G_{\rm light}} 
\newcommand{\gm}{G_{\rm matter}} 
\newcommand{\sige}{\sigma_8}

\begin{document}

\title{Constraining Cosmic Expansion and Gravity with Galaxy Redshift Surveys} 
\author{Scott F.\ Daniel$^{1}$ \& Eric V.\ Linder$^{2,3}$} 
\affiliation{$^1$Department of Astronomy, University of Washington, Seattle, 
WA, USA\\ 
$^2$Berkeley Center for Cosmological Physics \& Berkeley Lab, 
University of California, Berkeley, CA 94720, USA\\ 
$^3$Institute for the Early Universe WCU, Ewha Womans University, 
Seoul, Korea}

\begin{abstract}
We analyze the science 
reach of a next generation galaxy redshift survey such as BigBOSS to 
fit simultaneously for time varying dark energy equation of state and 
time- and scale-dependent gravity.  The simultaneous fit avoids potential 
bias from assuming $\Lambda$CDM expansion or general relativity and leads 
to only modest degradation in constraints.  Galaxy bias, fit freely in 
redshift bins, is self calibrated by spectroscopic measurements of 
redshift space distortions and causes little impact.  The combination 
of galaxy redshift, cosmic microwave background, and supernova distance 
data can deliver 5-10\% constraints on 6 model independent modified 
gravity quantities. 
\end{abstract}

\date{\today} 

\maketitle

%%%%%%%%%%%%%%%%%%%%%%%%%%%%%%%%%%%%%%%%%%%%%%%%%%%%%%%%%%%%%%%%%%%%%%%%
\section{Introduction} 

The power spectrum of matter density perturbations contains information on 
both the cosmic 
expansion history and growth history.  Comparison of these fundamental 
evolutions not only tightens constraints on the cosmological model but 
enables tests of the framework, such as the validity of general relativity 
as the theory of gravity.  However, most cosmological parameter estimation, 
from either current or projected data, either assumes general relativity, 
and hence fixes the growth history to be determined by the expansion history, 
or assumes a $\Lambda$CDM expansion history to test gravity. 

If either assumption is incorrect then the result will be biased.  Thus, 
even if one is only interested in constraining the expansion history and 
dark energy properties, for example, one still must fit for gravity in 
order to obtain robust results for the expansion.  Current data cannot 
tightly constrain either quantity without ad hoc assumptions such as 
no time dependence.  The volume and redshift reach of cosmological surveys 
is rapidly improving, however, and projections for the next generation 
are for the $\sim10\%$ accuracy level on the time variation of dark 
energy properties or gravity.

The question we address is how the constraints are affected when the 
matter power spectrum data measured by galaxy redshift surveys are fit 
simultaneously for varying dark energy 
equation of state, scale- and time-dependent gravity, and astrophysical 
evolution such as galaxy bias.  Aspects of this have been considered 
in the literature with some restrictions: for example \cite{stril} 
parametrized gravity with the scale- and time-independent gravitational 
growth index $\gamma$ and a single bias parameter, while \cite{fbias} 
extended this to binned bias; \cite{11094535,11094536} 
assumed scale-independent gravity with a particular time variation but 
allows scale- and time-dependent bias; \cite{zhao1109} applied a motivated 
scale- and time-dependent gravity parametrization to current data, 
restricting to quantities independent of galaxy bias.  (Note 
that more general simultaneous fitting, such as with principal component 
analysis, has been applied to other cosmological probes such as weak 
lensing, e.g.\ \cite{09051326,12103903}.) 

With the aim of deriving more general results that avoid assuming a 
particular model, for expansion history, gravity, or astrophysical bias, 
we use scale- and time-dependent bins of gravity and time varying effective 
dark energy equation of state and bins of galaxy bias.  Specifically, 
we fit for gravitational modifications to 
the two Poisson equations (for the behavior of matter and of light) as 
freely floating values at high and low redshift and large and small scales, 
i.e.\ in bins of redshift $z$ and wavenumber $k$.  Galaxy bias is fit 
as independent values in 17 bins of redshift.  Dark energy evolution is 
treated through the highly accurate $w_0$--$w_a$ parametrization.  
Due to the presence of non-Gaussian covariances between the many parameters 
we carry out a Markov Chain Monte Carlo exploration of parameter space 
using simulated next generation cosmological survey data. 

In Sec.~\ref{sec:model} we explain the treatment of dark energy, gravity, 
and galaxy bias in detail.  The simulated data sets used are presented in 
Sec.~\ref{sec:data}.  We analyze the results, with particular attention 
to covariances and the effect on constraints of not assuming fixed expansion, 
fixed gravity, or fixed galaxy bias in Sec.~\ref{sec:results}.  Implications 
for the science reach of next generation surveys are discussed in 
Sec.~\ref{sec:concl}.

%%%%%%%%%%%%%%%%%%%%%%%%%%%%%%%%%%%%%%%%%%%%%%%%%%%%%%%%% 
\section{Dark Energy, Gravity, and Galaxies} \label{sec:model} 

Matter density perturbations grow under gravitational instability but 
at a rate suppressed by cosmic expansion.  Therefore the evolution of 
large scale structure clustering depends on the matter density $\om$, 
the expansion rate $H$, and the laws of gravity.  In addition, galaxy 
redshift surveys do not directly measure the mass power spectrum in 
real space but the galaxy clustering in redshift space.  This introduces 
two additional ingredients: the statistical distribution of galaxies may 
be biased relative to the mass (i.e.\ dark matter), and the velocity 
field of the galaxies, generated by the gravitational potentials of 
the structures, adds anisotropic distortions to statistically isotropic 
density field. 

The redshift space distortion depends on the angle with respect to the 
line of sight and its overall amplitude is determined by the growth rate 
of structure at the given redshift, $f(z)$.  This carries with it important 
additional cosmological (and gravitational) information, enhancing the 
reach of spectroscopic galaxy surveys.  Galaxy bias $b$, however, has the 
potential to confuse extraction of the total amplitude of growth up to 
the given redshift.  Since it enters without angular dependence, one 
can fit separately the two effects, although each is convolved with the 
overall mass fluctuation amplitude often denoted by $\sige(z)$.  That is, 
cosmological information comes in the form of $b\sige(z)$ and $f\sige(z)$ 
\cite{percwhite}.  The amplitude $\sige(z)$ is proportional to the linear 
growth factor $D(z)$, which is related to the gravitational potential 
through a Poisson equation, and so this growth history depends on both the 
expansion history and gravity.  If additional data directly related to the 
mass fluctuations (rather than galaxies) is available, for example through 
weak gravitational lensing, then $D(z)$ can be separately determined, 
although lensing also depends on gravity, in its own way. 

Thus to draw general robust conclusions on expansion and dark energy 
properties, for example, we should fit not only for $H(z)$ but 
simultaneously for gravity -- the two histories then determining $D(z)$ and 
$f(z)=-d\ln D/d\ln(1+z)$ -- and galaxy bias.  Moreover, the way we treat 
each of these quantities should be valid over a wide range of cosmologies so 
that the results are not overly model dependent. 

For the expansion history we work within the flat Friedmann-Robertson-Walker 
framework, where the expansion rate is described by the matter density 
$\om$ as a fraction of the critical density and the time dependent dark 
energy equation of state $w(z)=w_0+w_a z/(1+z)$.  This form for $w(z)$, 
obtained from exact solutions for a wide range of dark energy and gravity 
models \cite{linprl}, has been demonstrated as a calibration relation with 
an accuracy of 0.1\% for observables \cite{calib}.  Perturbations in the 
effective dark energy (i.e.\ even if there is no physical dark energy, just 
a modification of gravity) are treated consistently within the equations of 
motion. 

Note that while many papers take an exact $\Lambda$CDM expansion history 
when allowing for modified gravity, this is highly model dependent.  While 
simple $f(R)$ gravity models can be treated in terms of the scalaron mass, 
which determines both the growth and expansion behaviors so that effective 
screening within the solar system (really, small $df/dR$) requires 
negligible deviations of $w(z)$ from $-1$, this is not generally true. 
For example, DGP gravity \cite{dgp} has strong deviations in both growth 
and expansion from $\Lambda$CDM, though as it is a one parameter model 
they are tied together.  Galileon models, however, have sufficient freedom 
that the significant deviations in expansion history can appear alongside 
moderate growth deviations (see, e.g., \cite{pathsgali}).  Thus fitting for 
expansion and gravity independently is most generic and free from possibly 
unwarranted assumptions. 

To account for possible extensions beyond general relativity, and their 
effects on matter growth and lensing of light, we solve for the two 
metric potentials $\phi$ and $\psi$ in modified Poisson equations 
\begin{eqnarray} 
\nabla^2 \psi&=&4\pi G_N a^2\delta\rho \times G_{\rm matter} \\ 
\nabla^2(\phi+\psi)&=&8\pi G_N a^2\delta\rho \times G_{\rm light} \ . 
\end{eqnarray} 
The rigor and completeness of these equations, together with those for 
the matter density and velocity from energy-momentum conservation, is 
discussed in detail in, e.g., \cite{daniel1002,10010969,10030001,daniel1008} 
(though here we have simplified the notation) and clearly related to the 
nonrelativistic and relativistic geodesic equations in \cite{edbert}. 

Parametrization of the generally time- and scale-dependent functions 
$\gm(k,z)$ and $\gl(k,z)$ can be specialized to forms expected from 
scalar-tensor or DGP gravity, for example (see \cite{zhao1109} for a 
unification of the two), or kept free form through a principal component 
analysis \cite{09051326,10030001,11113960,12103903}, say.  
Since we wish to carry out a substantially model independent analysis, 
but be able to interpret clearly physically the constraints on gravity, 
we follow the high vs.\ low redshift, large vs.\ small scale binning 
approach of \cite{daniel1002,daniel1008}.  This allows for 
8 free gravity parameters: $\gm$ and $\gl$ values in bins from $z=0$--1 
and $z=1$--2 (for $z>2$ their values are set to 1, i.e.\ behaving as 
general relativity in the high redshift, high curvature universe), and 
in wavenumber bins $10^{-4}{\rm Mpc}^{-1}<k<10^{-2}{\rm Mpc}^{-1}$ and 
$0.01\,{\rm Mpc}^{-1}<k<0.1\,{\rm Mpc}^{-1}$.  Because of the uncertainty in 
the appropriate growth equations in the nonlinear density region at higher 
wavenumbers for an arbitrary gravity theory (e.g.\ the presence of various 
screening mechanisms \cite{jainkhoury}), and in the velocity induced 
redshift space distortions to the galaxy power spectrum, we conservatively 
do not use data at $k>0.1\,{\rm Mpc}^{-1}$. 

Galaxy bias is treated through 17 free parameters representing the values 
$b(z)$ in bins of width 0.1 in redshift from $z=0.1$--1.8.  Since we only 
include large scale information we take the bias to be scale independent.  
We phrase the bias amplitude in terms of the comoving clustering expectation, 
$b(z)=b_0(z)\,D(z=0)/D(z)$ and fit for $b_0$ in each redshift bin.  
We use independent sets 
$\{b_{\rm LRG}(z_i)\}$ and $\{b_{\rm ELG}(z_i)\}$ for $b_0(z)$ for 
luminous red galaxies and emission line galaxies in the survey. 

In addition to these parameters we also fit for the physical baryon density 
$\Omega_b h^2$, physical matter density $\om h^2$, ratio of cosmic microwave 
background (CMB) sound horizon to angular diameter distance to last scattering 
$\theta$, optical depth to reionization $\tau$, scalar spectral tilt $n_s$, 
amplitude of primordial scalar perturbations $A_s$, and any astrophysical 
nuisance parameters appropriate to the data sets.

%%%%%%%%%%%%%%%%%%%%%%%%%%%%%%%%%%%%%%%%%%%%%%%%%%%%%%%%% 
\section{Galaxy Redshift Survey Science} \label{sec:data} 

Spectroscopic galaxy surveys provide three dimensional information on 
galaxy clustering and direct measurements of the galaxy density and 
velocity fields, thus probing the growth history.  BigBOSS \cite{bigboss} 
is planned to survey $50\,({\rm Gpc}/h)^3$ of cosmic volume, obtaining 
redshifts of 20 million galaxies from $z\approx0.5$--1.8.  
(In addition, it provides data 
on quasar clustering and on neutral hydrogen density fluctuations through 
the Lyman alpha forest, ranging over $z=1.8$--3.5, but conservatively we 
do not include constraints from those probes.)  

The redshift space galaxy power spectrum to be compared with observations 
is related to the linear theory mass power spectrum calculated from the 
equations of motion (Poisson equations and energy-momentum equations) by 
\be 
P_g^z({\mathbf{k}},z)=(b+f\mu^2)^2\,P_{m,lin}^r(k,z) \ , 
\ee 
where the squared factor is the Kaiser correction \cite{kaiser}.  This 
is a good approximation strictly in the linear regime; since we use only 
$k<0.1\,{\rm Mpc}^{-1}$ this is not too unreasonable.  How to map the 
isotropic mass power spectrum to the anisotropic galaxy power spectrum 
is not settled beyond linear theory so we do not attempt to go beyond the 
Kaiser formula here.  (See \cite{kll,fbias} for explorations of beyond 
linear theory constraints 
on the scale independent gravity parameter $\gamma$ as a function of 
maximum wavenumber, and \cite{jennfR} for specific $f(R)$ gravity models.) 

Uncertainties on the measurement of the galaxy power spectrum depend on 
the volume surveyed in each redshift shell and the number density $n$ of 
each type of galaxies used to sample the density field.  The volume is 
determined by the area of sky (14000 deg$^2$ for BigBOSS) and the 
cosmological distance-redshift relation.  We use the redshift distributions 
$n(z)$ for the two types of galaxies given in \cite{bigboss}.  See 
\cite{fkp,seoeis,fbias} for further details on calculating the likelihood 
with galaxy power spectrum measurements. 

In addition, we include simulated Planck CMB data.  This helps to 
constrain the cosmological parameters, especially the primordial ones, 
and the gravitational modifications through the integrated Sachs-Wolfe 
effect and CMB lensing.  The ISW effect has sensitivity to $\gl$ since 
this Poisson equation governs the behavior of relativistic geodesics; 
CMB lensing is sensitive to both $\gm$ through growth in the matter 
power spectrum and $\gl$ through light deflection.  
We include CMB lensing through the deflection angle power spectrum as
prescribed by \cite{Perotto:2006rj}. 
(See \cite{dawn} for its use in constraints on the gravity 
parameter $\gamma$.)  For constraints on the expansion history we employ 
simulated supernova distance-redshift measurements of 1800 supernovae 
over $z=0$--1.5 with a systematic floor of $0.02(1+z)/2.5$ magnitudes, 
and include the supernova absolute magnitude parameter $\mathcal{M}$ 
as a nuisance parameter. 

%%%%%%%%%%%%%%%%%%%%%%%%%%%%%%%%%%%%%%%%%%%%%%%%%%%%%%%%% 
\section{Constraining Expansion and Gravity Simultaneously} \label{sec:results} 
We carry out a Markov Chain Monte Carlo analysis to find the joint 
posterior likelihoods of the parameters, with our modified versions of 
CAMB \cite{camb} and CosmoMC \cite{cosmomc}.  Our baseline analysis 
consists of projected galaxy clustering, CMB, and supernova data, allowing 
for independent fits of the 8 binned gravity parameters $\gl$ and $\gm$, 
27 binned galaxy bias parameters for LRG and ELG, 
the dark energy equation of state parameters $w_0$, $w_a$, plus the other 
cosmological and nuisance parameters.  
To speed the convergence of our Markov Chain Monte Carlo implementation,
we fit for the maximum-likelihood bias combination at each step in cosmological
parameter space.  Because of this, our quoted credibility regions should be
considered conservative, as outlying points in cosmological parameter space are
assigned a larger posterior density than they might be if we marginalized over
the bias parameters.
We then consider the following 
variations to explore the impacts of physical ingredients: 

\begin{itemize} 
\item Expansion: fix to $\Lambda$CDM background, i.e.\ $w_0=-1$, $w_a=0$ 
\item Gravity: fix to general relativity, i.e.\ $\gm=1$, $\gl=1$ for all bins 
\item Galaxy bias: fix to inverse growth, but fit for constant 
$b_0^{\rm LRG}$, $b_0^{\rm ELG}$ 
\end{itemize} 

\noindent We emphasize that our baseline case varies all these ingredients. 

\begin{figure*}[!t] 
\includegraphics[width=\columnwidth]{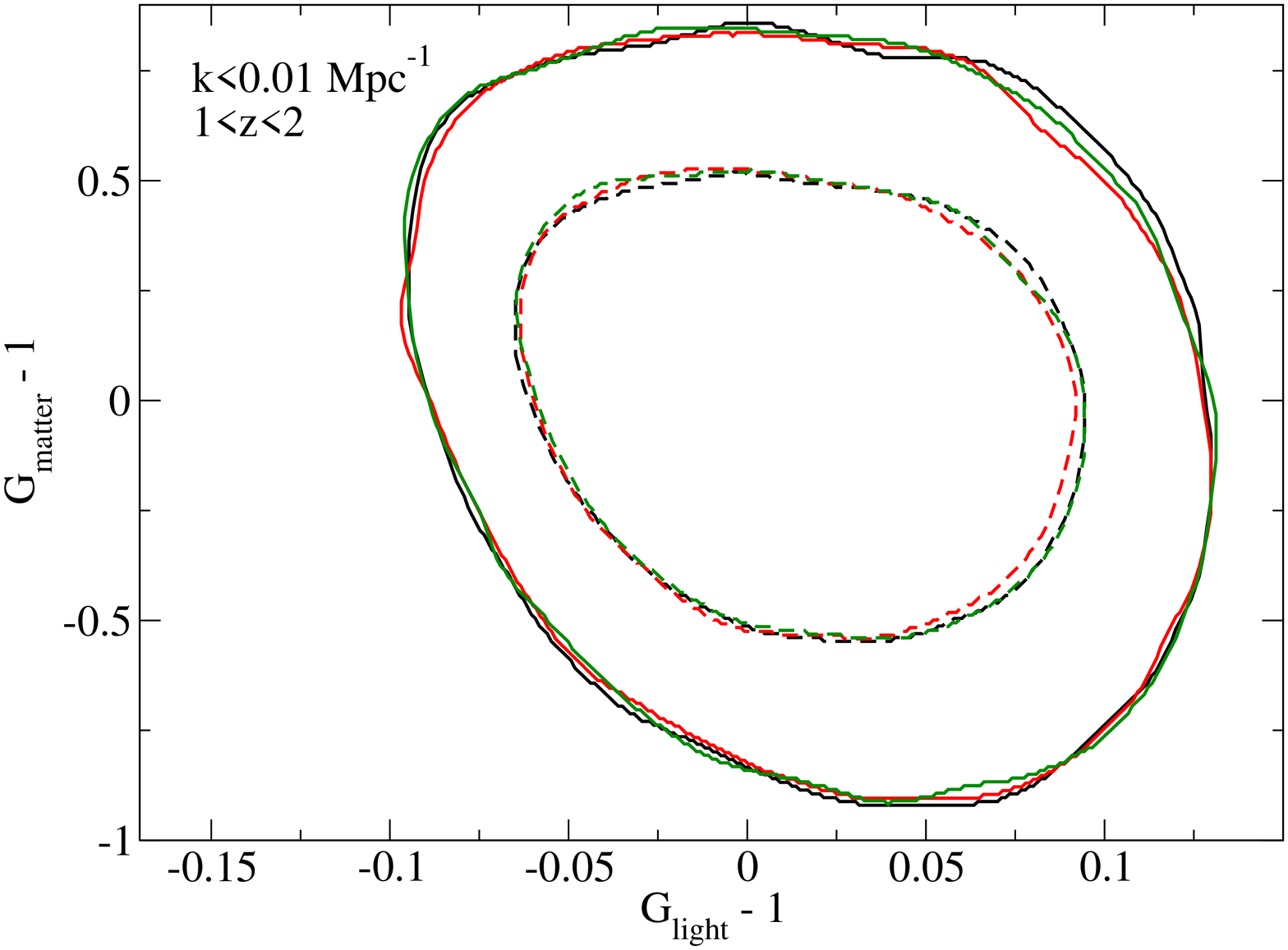}
\includegraphics[width=\columnwidth]{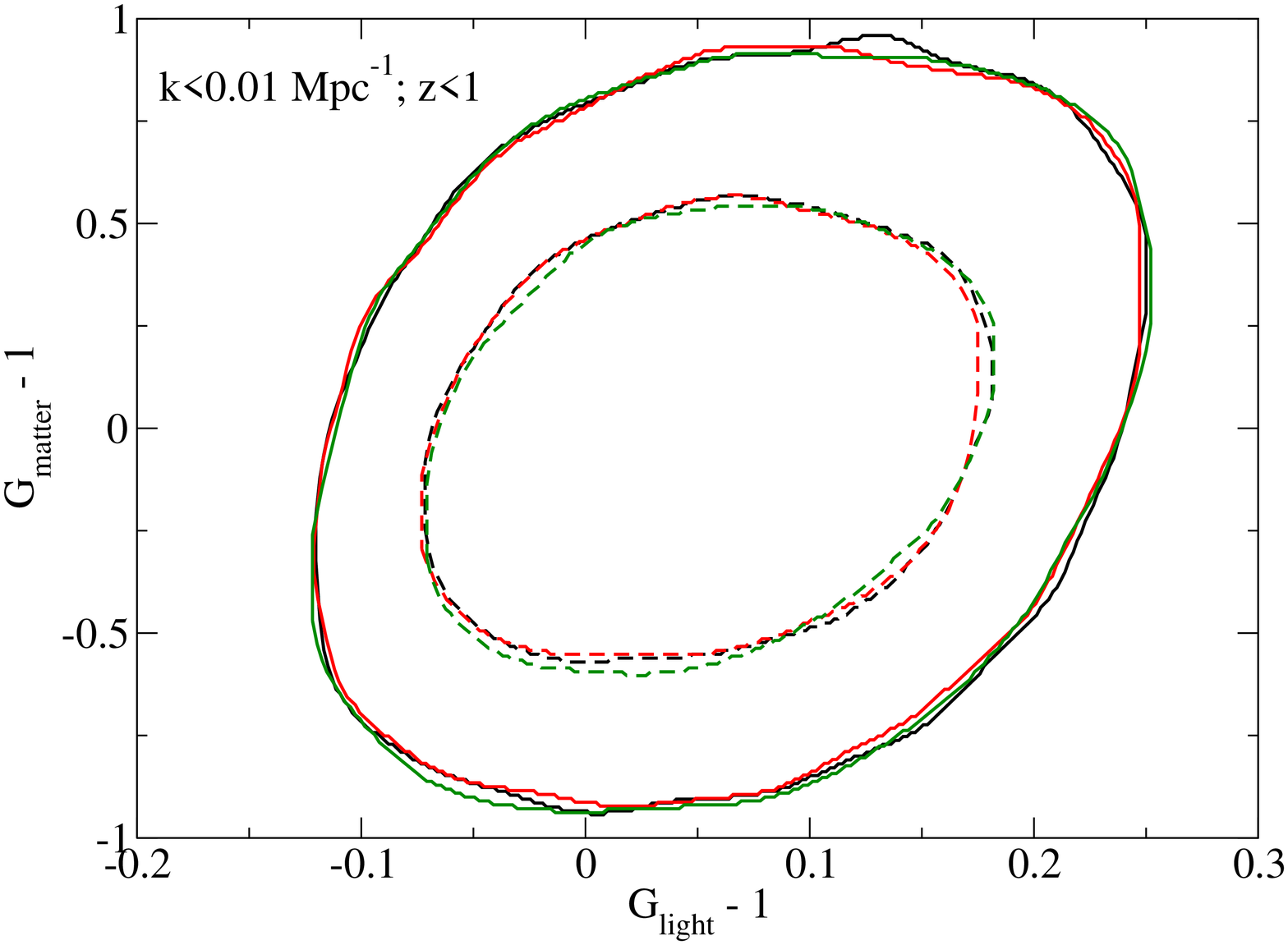}
\includegraphics[width=\columnwidth]{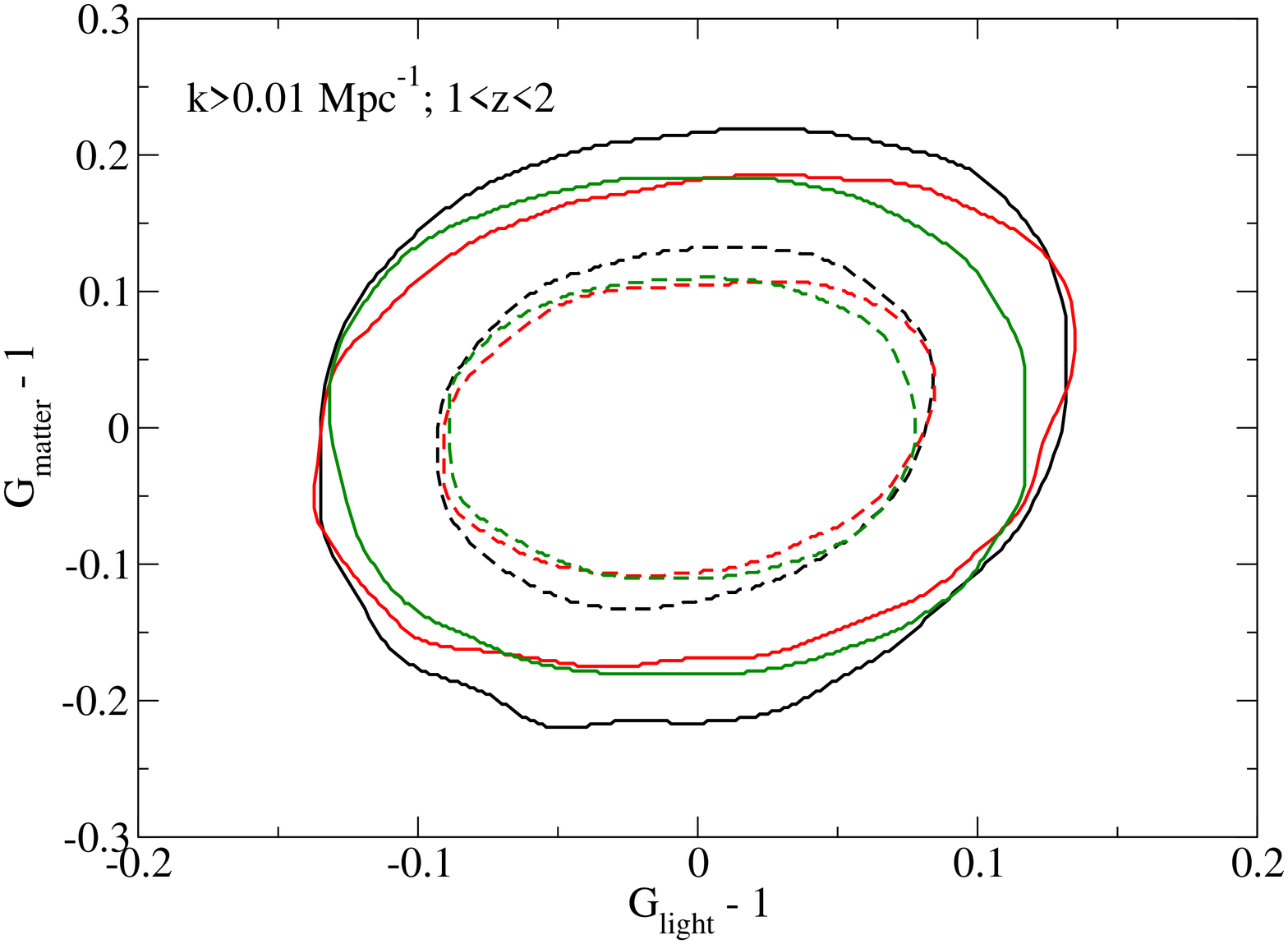}
\includegraphics[width=\columnwidth]{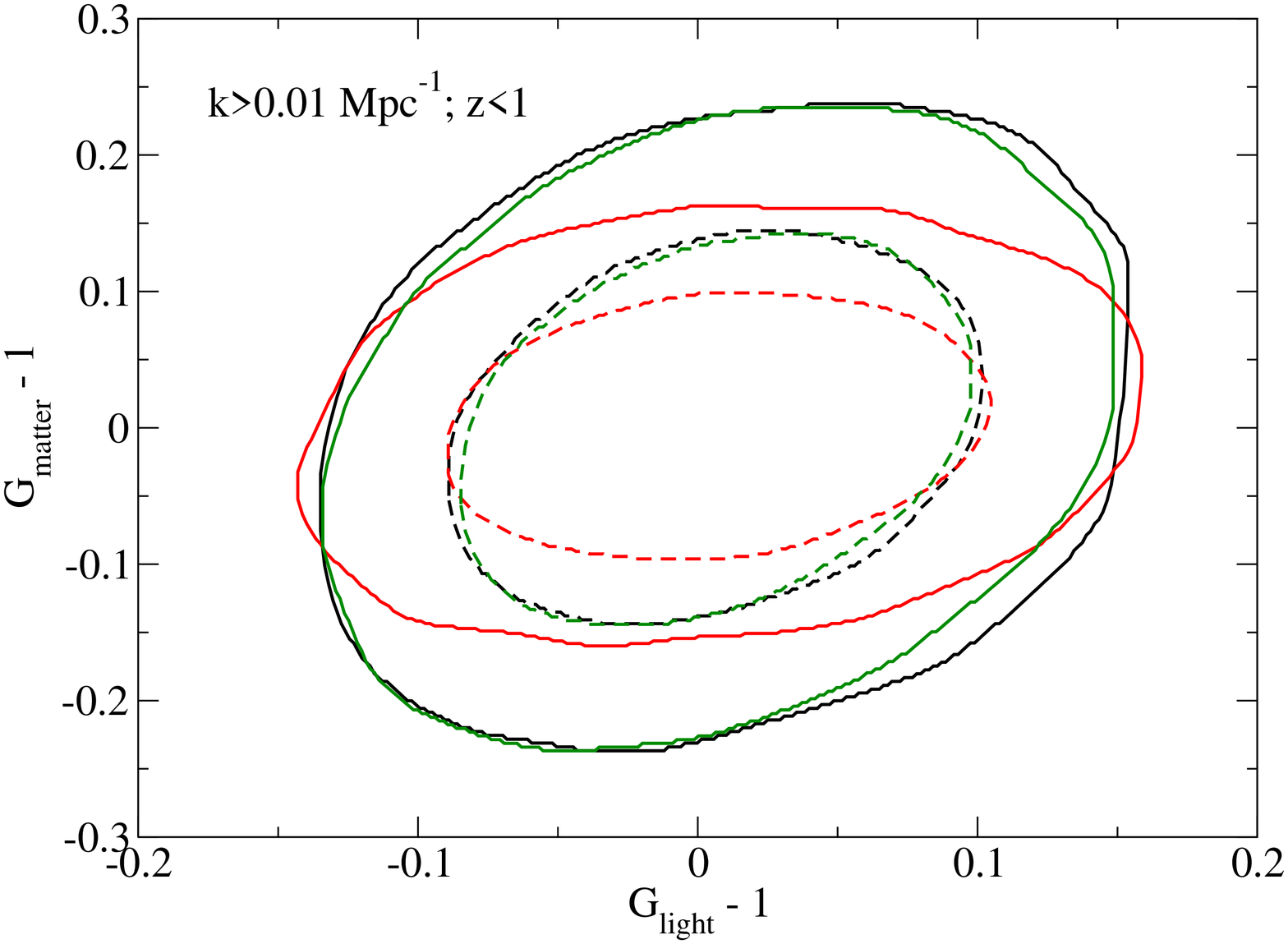} 
\caption{68\% and 95\% projected confidence limit contours for the 
gravitational 
parameters binned in low and high redshift, low and high $k$ are plotted 
for the baseline case of fitting for gravity, expansion, and galaxy bias 
(black curves), and the variations of fixing expansion by $w=-1$ (light 
green curves) or fixing galaxy bias to inverse growth (dark red curves).  
} 
\label{fig:gg} 
\end{figure*}

The results for the gravitational sector in the baseline case and its 
physical variations are shown in Figs.~\ref{fig:gg}.  We see that for 
next generation data sets there is little degradation in fitting for 
dark energy equation of state and galaxy bias simultaneously with gravity, 
giving great promise to our ability to test the cosmological framework. 
Adding parameters for non-$\Lambda$CDM expansion has only a percent level 
effect (except $\sim15\%$ on $\gm$ at large $k$ and $z$), 
while bias parameters only affect $\gm$ at large $k$ and only loosen 
the constraints by at most $\sim50\%$.  

The quantity $\gl$ will be determined to $\sim5\%$ (at 68\% CL) 
independently at low and high redshift, and small and large wavenumber, 
while $\gm$ will be tested to $\sim8\%$ on small scales 
($k>0.01 \text{Mpc}^{-1}$) where the galaxy survey data is most important, 
and to $\sim35\%$ on large scales ($k<0.01 \text{Mpc}^{-1}$).  
The much weaker constraints at large scales, where we assume no 
galaxy data, demonstrate the importance of the galaxy power spectrum 
measurement in constraining gravity.  
Even so, the constraints on large scale $\gm$ show more than a factor of 
two improvement over the yellow shaded contours in Figure~11 of 
\cite{daniel1008} (their $\mathcal{V}$ is our $\gm$), due to our inclusion 
of CMB lensing.  

The gravity parameters have little covariance with each other (including 
between large and small scale bins of the same parameter, not shown), except 
mildly at low redshift.  Excellent complementarity exists between the 
quantities and the probes: galaxy data constrains $\gm$, CMB data 
predominantly constrains $\gl$ (again see Figure~11 of \cite{daniel1008} 
and Figure~7 of \cite{daniel1002}), and supernova data constrains the 
expansion dynamics of $w_0$, $w_a$, ignoring the gravity parameters.

%%%%%%%%%%%%%%%%%%%%%%%%%%%

\begin{figure} 
\includegraphics[width=\columnwidth]{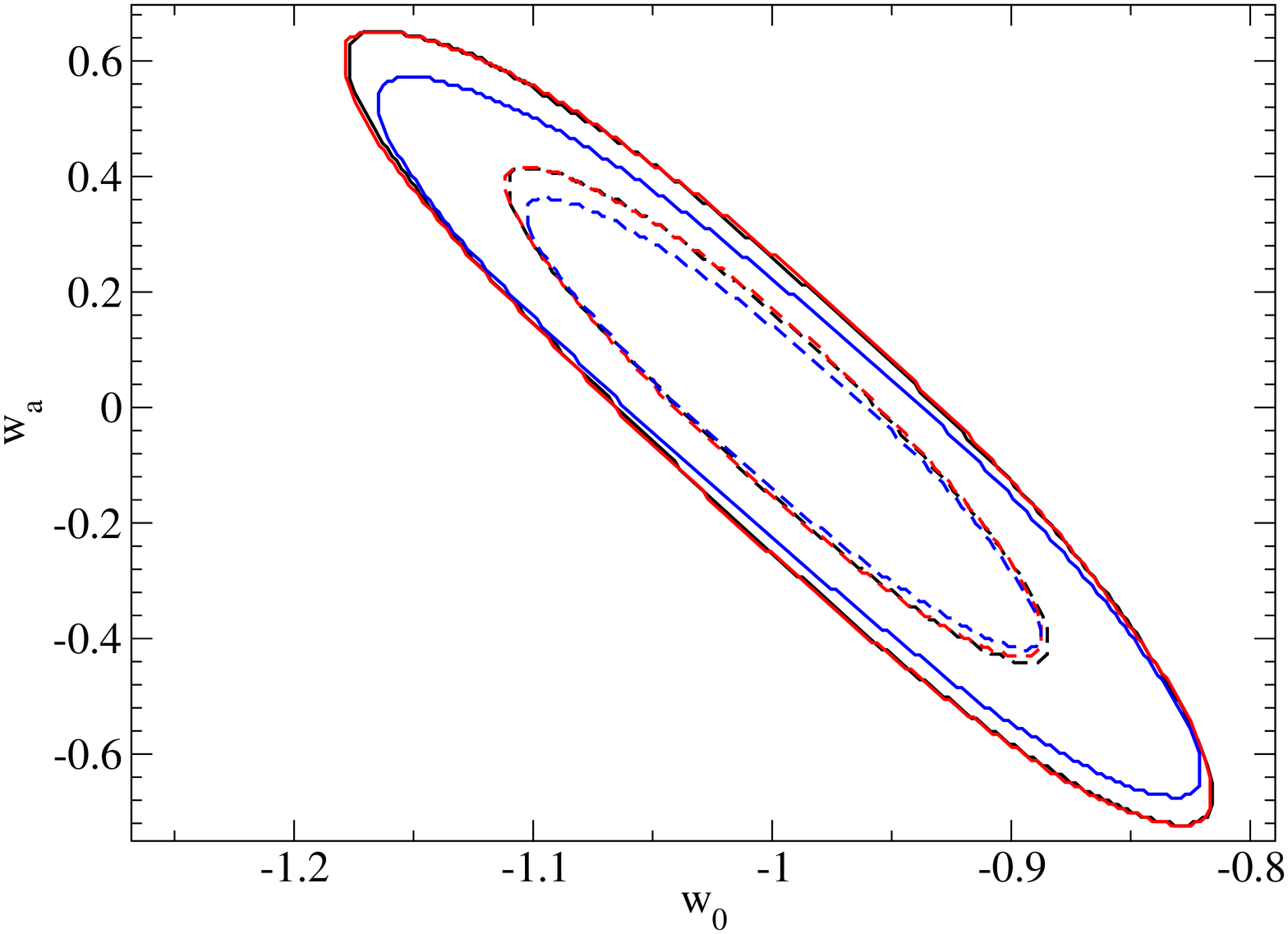}
\caption{68\% and 95\% confidence limit contours for the effective dark 
energy equation of state parameters $w_0$ and $w_a$ are plotted for the 
baseline case of fitting for gravity, expansion, and galaxy bias
(black curves), and the variations of fixing to general relativity 
(light blue curves) or fixing galaxy bias to inverse growth (dark red 
curves). 
} 
\label{fig:ww} 
\end{figure}

The expansion sector is also not strongly affected by the different 
cases, when next generation data is available.  Figure~\ref{fig:ww} 
shows the contours in the baseline case, when fixing galaxy bias, and 
when fixing to general relativity.  The $w_0$--$w_a$ contour increases 
in area by $\sim30\%$ when including the fit for the 8 gravity parameters, 
with $\sim2\%$ loosening in $w_0$ and $\sim20\%$ in $w_a$ determination. 
As expected from these and the previous results, plots of the 
crosscorrelations of gravity and expansion parameters (not shown) display 
little covariance, 
with only a minor correlation between $w_a$ and $\gm$ at large $k$.

%%%%%%%%%%%%%%%%%%%%%%%%%%%%%%%%% 

%%%%%%%%%%%%%%%%%%%%%%%%%%%%%%%%%%%%%%%%%%%%%%%%%%%%%%%%% 
\section{Conclusions} \label{sec:concl} 

Testing gravity on cosmic scales is crucial to understanding acceleration 
and fundamental laws of physics.  
The inability of present data sets to significantly constrain the laws 
of gravity affecting matter (i.e.\ $\gm$) is the largest obstacle to 
efforts to test general relativity on cosmological scales and to constrain 
modified gravity theories.  Next generation astronomical
surveys propose to remove this obstacle by presenting us with detailed
redshift-space maps of the distributions of galaxies throughout the 
Universe.  

We have shown that these maps will, indeed, be able to deliver on their 
promise to constrain gravity at cosmological scales, even given our 
imperfect knowledge of both the cosmic expansion history and the bias 
factor relating galaxy to dark matter clustering.  While a theoretical 
framework predicting the galaxy-dark matter bias would certainly be 
welcome, the results show it is not requisite for cosmological tests of 
gravity.  Physics such as a generalized expansion history 
and astrophysics such as galaxy bias should be included in the fits 
(rather than assumed known) to avoid erroneous 
conclusions about gravity, and can be included without fear of degrading 
observational constraints beyond usefulness.  Conversely, future experiments 
can proceed without fear of their science output being obscured by our 
ignorance. 

The model independent approach to testing gravity used here avoids 
restriction to a specific model, and has excellent orthogonality among 
its quantities and between its quantities and the expansion and galaxy 
bias parameters.  Strong complementarity exists as well between the 
cosmic probes: galaxy surveys testing $\gm$, CMB measurements constraining 
$\gl$, and supernova distances measuring expansion.  Together, the 
next generation combination of these data will deliver 5--10\% measurements 
of 6 gravity quantities, plus 35\% measures of the remaining 2.  These can 
potentially be improved further by going beyond the purely linear regime 
(we conservatively cut off $k>0.1\,{\rm Mpc}^{-1}$) once modified gravity 
effects on the redshift-space galaxy power spectrum are better understood, 
by higher quality CMB lensing from ground based polarization experiments, 
or by inclusion of crosscorrelations between galaxy maps and the CMB.

%%%%%%%%%%%%%%%%%%%%%%%%%%%%%%%%%%%%%%%%%%%%%%%%%%%%%%%%%%%%    
\acknowledgments

This work has been supported by World Class University grant 
R32-2009-000-10130-0 through the National Research Foundation, Ministry 
of Education, Science and Technology of Korea and DOE grant DE-SC-0007867 
and the Director, 
Office of Science, Office of High Energy Physics, of the U.S.\ Department 
of Energy under Contract No.\ DE-AC02-05CH11231. 
SFD acknowledges support from DOE award grant number DE-SC-0002607.

%%%%%%%%%%%%%%%%%%%%%%%%%%%%%%%%%%%%%%%%%%%%%%%%%%%%%%%%%%%%


\begin{thebibliography}{}

\bibitem{stril} 
A. Stril, R.N. Cahn, E.V. Linder, MNRAS 404, 239 (2010) [arXiv:0910.1833] 

\bibitem{fbias} 
E.V. Linder \& J. Samsing, arXiv:1211.2274 

\bibitem{11094535} 
I. Laszlo, R. Bean, D. Kirk, S. Bridle, MNRAS 423, 1750 (2012) 
[arXiv:1109.4535] 

\bibitem{11094536} 
D. Kirk, I. Laszlo, S. Bridle, R. Bean, arXiv:1109.4536 

\bibitem{zhao1109} 
G-B. Zhao, H. Li, E.V. Linder, K. Koyama, D.J. Bacon, X. Zhang, 
Phys. Rev. D 85, 123546 (2012) [arXiv:1109.1846] 

\bibitem{09051326} 
G-B. Zhao, L. Pogosian, A. Silvestri, J. Zylberberg, Phys. Rev. Lett. 103, 
241301 (2009) [arXiv:0905.1326] 

\bibitem{12103903} 
A. Hojjati, arXiv:1210.3903 

\bibitem{percwhite} 
W.J. Percival \& M. White, MNRAS 393, 297 (2009) [arXiv:0808.0003] 

\bibitem{linprl} 
E.V. Linder, Phys. Rev. Lett. 90, 091301 (2003) [arXiv:astro-ph/0208512] 

\bibitem{calib} 
R. de Putter \& E.V. Linder, JCAP 0810, 042 (2008) 
[arXiv:0808.0189] 

\bibitem{dgp} 
G. R. Dvali, G. Gabadadze, M. Porrati, Phys. Lett. B 485, 208 (2000) 
[arXiv:hep-th/0005016] 

\bibitem{pathsgali} 
S.A. Appleby \& E.V. Linder, JCAP 1203, 043 (2012) [arXiv:1112.1981] 

\bibitem{daniel1002} 
S.F. Daniel, E.V. Linder, T.L. Smith, R.R. Caldwell, A. Cooray,  
A. Leauthaud, L. Lombriser, Phys. Rev. D 81, 123508 (2010) [arXiv:1002.1962] 

\bibitem{10010969} 
Y-S. Song, L. Hollenstein, G. Caldera-Cabral, K. Koyama, JCAP 1004, 018 
(2010) [arXiv:1001.0969] 

\bibitem{10030001} 
G.B. Zhao et al., Phys. Rev. D 81, 103510 (2010) [arXiv:1003.0001] 

\bibitem{daniel1008} 
S.F. Daniel \& E.V. Linder, Phys. Rev. D 82, 103523 (2010) [arXiv:1008.0397] 

\bibitem{edbert} 
E. Bertschinger \& P. Zukin, Phys. Rev. D 78, 024015 (2008) [arXiv:0801.2415] 

\bibitem{11113960} 
A. Hojjati, G-B. Zhao, L. Pogosian, A. Silvestri, R. Crittenden, K. Koyama, 
Phys. Rev. D 85, 043508 (2012) [arXiv:1111.3960] 

\bibitem{jainkhoury} 
B. Jain \& J. Khoury, Annals Phys. 325, 1479 (2010) [arXiv:1004.3294] 

\bibitem{bigboss} 
D. Schlegel et al, arXiv:1106.1706 ; \url{http://bigboss.lbl.gov} 

\bibitem{kaiser} 
N. Kaiser, MNRAS 227, 1 (1987) 

\bibitem{kll} 
J. Kwan, G.F. Lewis, E.V. Linder, ApJ 748, 78 (2012) [arXiv:1105.1194] 

\bibitem{jennfR} 
E. Jennings, C.M. Baugh, B. Li, G-B. Zhao, K. Koyama, MNRAS 425, 2128 (2012) 
[arXiv:1205.2698] 

\bibitem{fkp} 
H.A. Feldman, N. Kaiser, J.A. Peacock, ApJ 426, 23 (1994) 
[arXiv:astro-ph/9304022] 

\bibitem{seoeis} 
H-J. Seo \& D.J. Eisenstein, ApJ 598, 720 (2003) [arXiv:astro-ph/0307460] 

\bibitem{Perotto:2006rj}
L. Perotto, J. Lesgourgues, S. Hannestad, H. Tu, Y.Y.Y. Wong,
JCAP 0610, 013 (2006) [arXiv:astro-ph/0606227]

\bibitem{dawn} 
S. Das \& E.V. Linder, Phys. Rev. D 86, 063520 (2012) [arXiv:1207.1105] 

\bibitem{camb} 
A. Lewis, A. Challinor, A. Lasenby, ApJ 538, 473 (2000) 
[arXiv:astro-ph/9911177]; \url{http://camb.info} 

\bibitem{cosmomc} 
A. Lewis \& S. Bridle, Phys. Rev. D 66, 103511 (2002) 
[arXiv:astro-ph/0205436] \\ 
\url{http://cosmologist.info/cosmomc} 


\end{thebibliography}
\end{document}